# Tunable Topological Dirac Surface States and Van Hove Singularities in Kagome Metal GdV$_6$Sn$_6$

Yong Hu[1,#,*], Xianxin Wu[2,3,#], Yongqi Yang[4], Shunye Gao[1,5], Nicholas C. Plumb[1], Andreas P. Schnyder[3], Weiwei Xie[4], Junzhang Ma[6,7,8,*], Ming Shi[1,*]

[1]Swiss Light Source, Paul Scherrer Institut, CH-5232 Villigen PSI, Switzerland
[2]CAS Key Laboratory of Theoretical Physics, Institute of Theoretical Physics, Chinese Academy of Sciences, Beijing 100190, China
[3]Max-Planck-Institut für Festkörperforschung, Heisenbergstrasse 1, D-70569 Stuttgart, Germany
[4]Department of Chemistry and Chemical Biology, Rutgers University, 08854, Piscataway, USA
[5]Beijing National Laboratory for Condensed Matter Physics and Institute of Physics, Chinese Academy of Sciences, Beijing, 100190, China
[6]Department of Physics, City University of Hong Kong, Kowloon, Hong Kong, China
[7]City University of Hong Kong Shenzhen Research Institute, Shenzhen, China
[8]Hong Kong Institute for Advanced Study, City University of Hong Kong, Kowloon, Hong Kong, China

#These authors contributed equally to this work.
*To whom correspondence should be addressed:
Y.H. (yonghphysics@gmail.com); J.Z.M. (junzhama@cityu.edu.hk); M.S. (ming.shi@psi.ch)

**Transition-metal-based kagome materials at van Hove filling are a rich frontier for the investigation of novel topological electronic states and correlated phenomena. To date, in the idealized two-dimensional kagome lattice, topologically nontrivial Dirac surface states (TDSSs) have not been unambiguously observed, and the manipulation of TDSSs and van Hove singularities (VHSs) remains largely unexplored. Here, we reveal TDSSs originating from a $\mathbb{Z}_2$ bulk topology and identify multiple VHSs near the Fermi level ($E_F$) in magnetic kagome material GdV$_6$Sn$_6$. Using *in-situ* surface potassium deposition, we successfully realize manipulation of the TDSSs and VHSs. The Dirac point of the TDSSs can be tuned from above to below $E_F$, which reverses the chirality of the spin texture at the Fermi surface. These results establish GdV$_6$Sn$_6$ as a fascinating platform for studying the nontrivial topology, magnetism and correlation effects native to kagome lattices. They also suggest potential application of spintronic devices based on kagome materials.**

**INTRODUCTION**

Realizing and tuning novel electronic states is of great interest and importance to modern condensed-matter physics and spintronics applications. The exploration of topological physics intertwined with nontrivial lattice geometries and strong electron interactions is emerging as a new frontier in condensed-matter physics (1-27). Transition-metal-based kagome lattices, owing to the unique lattice geometry, have attracted particular attention, as they often show correlated topological band structures, magnetism, and diverse exotic electronic instabilities, such as spin liquid states, charge density wave (CDW), and superconductivity (1-9). Recently, a substantial number of experimental efforts have investigated topological phenomena in transition-metal-based kagome magnets, such as $Fe_3Sn_2$ (3), FeSn (7), $Co_3Sn_2S_2$ (4,11,12), and $RMn_6Sn_6$ (where R is a rare-earth element, and the kagome layers are made of manganese atoms) (8,13-15). In particular, the topological surface states derived from relativistic Weyl points, namely surface Fermi arcs, have been observed in the time-reversal symmetry-broken Weyl semimetal $Co_3Sn_2S_2$ with ferromagnetism (11,12). In addition, various forms of magnetism in the low-temperature electronic ground state enable the realization of quantum-limit Chern topological magnetism (8). Besides these intensively studied magnetic kagome materials, the recently discovered non-magnetic kagome metals $AV_3Sb_5$ (A=K, Rb, Cs) were found to possess a unique combination of novel correlated phenomena and nontrivial band topology (9,10,16-20). An intriguing chiral CDW instability (21,22) and superconductivity (10,16,17) have been observed in the absence of magnetic orders, suggesting that the vanadium kagome lattice is an ideal platform for investigating correlated quantum states. Moreover, Dirac nodal lines and nodal loops have been identified in the non-magnetic kagome metal $CsV_3Sb_5$ (23). Apart from various types of topological band crossings (e.g., Dirac and Weyl fermions), $\mathbb{Z}_2$ topology in kagome materials has also been proposed by theoretical calculations (16); however, identifying and confirming the corresponding topological nature of surface states have remained an outstanding challenge, due to the lack of good candidate systems. Therefore, no clear experimental evidence for the $\mathbb{Z}_2$ topological Dirac surface states (TDSSs) in kagome materials has been reported to date. For instance, the expected topological surface states (TSSs) in $CsV_3Sb_5$ from density functional theory (DFT) calculations lie above the Fermi level ($E_F$) and mostly overlap with the projection of bulk states (16,24). Moreover, it has remained unclear how these kagaome surface states could be manipulated, a key question relevant for potential applications.

GdV$_6$Sn$_6$ is a newly discovered kagome system that exhibits a magnetic transition at a low temperature $T_m$ ~ 5 $K$ (25,26). In contrast to other members of the kagome magnet family, it is the Gd-triangular lattice that generates magnetism, while the kagome layer composed of V and Sn atoms is non-magnetic (Fig. 1A). The separation of the magnetic layer and kagome layer not only permits a direct study of the electronic structure of non-magnetic kagome layer but also introduces a magnetic tunability from the magnetic layer below $T_m$. Band structure calculations suggest that GdV$_6$Sn$_6$ is topologically nontrivial, characterized by a $\mathbb{Z}_2$ topological invariant in the paramagnetic state (25). In contrast to other $\mathbb{Z}_2$ kagome metals, such as CsV$_3$Sb$_5$, GdV$_6$Sn$_6$ has a large bulk gap around $\Gamma$, allowing the TDSSs to be well separated from bulk states around the surface Brillouin zone (BZ) center ($\bar{\Gamma}$ point) (as illustrated in Fig. 1B). Therefore, GdV$_6$Sn$_6$ is a tantalizing system to access and tune TDSSs, which is crucial for exploring potential applications in spintronics. Moreover, multiple van Hove singularities (VHSs), originating from the vanadium $d$ orbitals appear near the $E_F$ at the $\bar{M}$ point (Fig. 1C), providing a promising playground in the search for exotic correlated states on the kagome lattice. While theory predicts the nontrivial band topology and the great potential for nesting effects around the $\bar{M}$ points at van Hove filling in GdV$_6$Sn$_6$, the existence of the TDSSs – as well as the manipulation of the TDSSs and VHSs – in kagome metals have yet to be experimentally demonstrated.

In this work, via a combination of angle-resolved photoemission spectroscopy (ARPES) and DFT calculations, we unambiguously reveal the characteristic $\mathbb{Z}_2$ TDSSs in kagome lattices and identify two types ($p$-type and $m$-type) of VHSs at the M points, in the paramagnetic phase of the magnetic kagome metal GdV$_6$Sn$_6$. The direct manipulation of the TDSSs and VHSs is realized by surface potassium deposition, where the Dirac point of the TDSSs shifts from above to below the $E_F$ with increasing electron doping. The direct identification of surface states, together with the spin texture inferred from spin-resolved ARPES (spin-ARPES) measurements and theoretical calculations, confirms the bulk nontrivial $\mathbb{Z}_2$ topology and shows great promise for realizing spin polarization reversal on the surface Fermi surfaces in GdV$_6$Sn$_6$. Our observation of tunable correlated and topological electronic states not only establishes GdV$_6$Sn$_6$ as a fertile system for exploring the interplay between the nontrivial band topology, magnetism, and correlation effects native to kagome lattices, but also unlocks new perspectives for the realization of spintronics devices based on kagome materials.

**RESULTS**

GdV$_6$Sn$_6$ has a layered crystal structure with the space group P6/mmm and hexagonal lattice constants a = 5.5 Å and c = 9.2 Å. It consists of V$_3$Sn kagome layers with Sn and GdSn$_2$ layers successively distributed in alternating layers stacked along the *c* axis [Fig. 1A, (i) and (ii)] (25). From the crystal structure, we find that chemical bonding between V$_3$Sn and Sn layers is strong while the bonding between the V$_3$Sn and GdSn$_2$ layers is weaker. Therefore, cleaving the crystal along (001) direction will result in two possible surface terminations, namely the V kagome and Gd terminations [marked as Kagome Term. and Gd Term. in Fig. 1A (iii)]. Figure 1D illustrates the bulk BZ and the projected two-dimensional (001) surface BZ, with high-symmetry points indicated. The band structure of GdV$_6$Sn$_6$ in the paramagnetic phase from DFT calculations is displayed in Fig. 1E, where four VHS points emerge at the M in the vicinity of $E_F$ (indicated by the red arrows and labeled as VHS1-4). A closer examination of the orbitally decomposed electronic structure from DFT indicates that the states of VHS1, VHS2 and VHS3 at the M point, characterized by V $d_{xz}$, $d_{xy}$ and $d_{z2}$ orbitals, are solely attributed to one sublattice in the V kagome lattice, and thus are of *p*-type (19,20,27). In contrast, the states of VHS4 at the M point are attributed to a mixture of two sublattices, and thus are of *m*-type (for details, see fig. S1). Interestingly, along the K-M direction, the *p*-type VHS1 and VHS2 bands disperse with the opposite sign compared to the VHS3 bands, which is attributed to the sign change of the hopping parameters of $d_{xz}$ and $d_{xy}$ orbitals. The continuous direct gaps between bands appearing at every *k* point allow one to define the $\mathbb{Z}_2$ topological invariant for the occupied bands using parity products at time-reversal invariant momenta (28). Consistent with previous calculations (25), a strong topological invariant $\mathbb{Z}_2$ = 1 is assigned to bands 171 (black) and 169 (blue), while the topmost band 173 (red) is topologically trivial. Owing to the band inversion around the A point for the 172 occupied bands, Dirac-cone-like TSSs are expected to reside in the large local band gap at $\Gamma$ (fig. S2). Motived by these theoretical observations, we employ ARPES to systematically study the topological electronic structures of single-crystal GdV$_6$Sn$_6$.

Due to the two possible surface terminations upon cleaving [Fig. 1A (iii)], we expect to observe two different types of ARPES spectra (29). By using a small beam spot and measuring the Sn 4*d* core level, we have resolved the two types of terminations on the cleaved sample surface and probed their electronic structures separately. Figure 2 summarizes the photoemission experiments on pristine, freshly cleaved GdV$_6$Sn$_6$.

We first focus on the electronic structure from the kagome termination (see fig. S3 for a detailed description of the termination assignment). The corresponding X-ray photoelectron spectroscopy (XPS) spectrum on the Sn 4$d$ core level and Fermi surface (FS) are shown in Fig. 2 (A and B), respectively (also see fig. S3). Photon energy-dependent ARPES measurements on the kagome V layer along two different high-symmetry paths, i.e., the $\bar{\Gamma}$ - $\bar{K}$ - $\bar{M}$ - $\bar{K}$ (compare Figs. 2C and 2E) and $\bar{\Gamma}$ - $\bar{M}$ - $\bar{\Gamma}$ (Fig. 2, D and F) directions, exhibit distinct band dispersions at different $k_z$ planes, indicating the three-dimensionality of the electronic structure in GdV$_6$Sn$_6$ (consistent with the calculations in Fig. 1E). Similar to other kagome lattices, the characteristic Dirac cone around the $\bar{K}$ point and the VHS point near $\bar{M}$ of the kagome lattice are observed on the kagome termination (fig. S4).

The ARPES spectra collected on the Gd termination, shown in Fig. 2 (G to J), are even richer than that on the kagome termination. The XPS spectrum on the Sn core level and FS from the Gd termination are plotted in Fig. 2 (G and H), respectively. The most prominent features of the FS (Fig. 2H) are the circular-shaped pocket near the BZ center ($\bar{\Gamma}$ point, highlighted by the dashed circle in Fig. 2H and the accompanying hexagonal-shaped sheet (dashed hexagon in Fig. 2H). The band dispersion across the $\bar{\Gamma}$ point (Fig. 2J) further reveals that the circular- and hexagonal-shaped Fermi surfaces are formed by two V-shaped bands (highlighted by the red box in Fig. 2J). In addition, the measured dispersions uncover an electron-like band (centered at the $\bar{M}$) along the $\bar{\Gamma}$ - $\bar{K}$ - $\bar{M}$ - $\bar{K}$ direction (Fig. 2I) and a hole-like band along the $\bar{\Gamma}$ - $\bar{M}$ - $\bar{\Gamma}$ direction (Fig. 2J), exhibiting a saddle point at the $\bar{M}$ point (Fig. 2K), i.e., indicating the presence of a VHS, as sketched in Fig. 1C. A careful comparison (fig. S5) between the experimental and calculated bands shows good overall agreement, and indicates that the identified VHS corresponds to the VHS1 labeled in Fig. 1E. On the other hand, the observed double V-shaped bands around the $\bar{\Gamma}$ point appear within the bulk band gap (Fig. 1E), implying the presence of surface states on the kagome metals GdV$_6$Sn$_6$.

To further validate the surface nature of double V-shaped bands around the $\bar{\Gamma}$ point, we have conducted photon energy-dependent ARPES measurement. Photoemission spectra recorded at various photon energies from 40 to 100 $eV$ reveal that the V-shaped bands around $\bar{\Gamma}$ do not disperse with respect to photon energy (and thus $k_z$), in contrast to the bulk states (for details, see fig. S6). The $k_z$ independence of the V-shaped bands, as illustrated by two representative spectra taken with 76 $eV$ and 86 $eV$ in Fig. 3A, confirms the two-dimensional nature of the discussed surface bands. To further explore the topological nature of the surface states, we plot in Fig. 3 (B and C) the calculated bulk states projected onto the (001) surface together with the theoretical surface spectra for the Gd termination (also see fig. S7). Comparing the measurements (Fig. 3A) with the theory calculations (Fig. 3, B and C), the TSSs derived from bulk nontrivial topology are identified around the $\bar{\Gamma}$ point (indicated by the black arrows in Fig. 3C). Notably, the side-by-side comparisons of the band dispersion along the $\bar{\Gamma}$ - $\bar{M}$ direction (Figs. 3A and 3D) and the evolution of constant energy contours at different binding energies (Fig. 3, E and F) show excellent agreement. Moreover, spin-ARPES results presented in Fig. 3G provide clear spectroscopic evidence of the spin character of the V-shaped bands, suggesting that the observed surface states originate from the bulk nontrivial $\mathbb{Z}_2$ topology in kagome metals (for details see fig. S8).

After the identification of the TSSs and VHS bands, we further demonstrate their manipulation via *in situ* surface potassium deposition. The accumulation of the K atoms on the sample surface can be seen by measuring the K 3*p* core level, which is absent on the pristine surface (blue curve in Fig. 4A), but grows in intensity with the surface deposition (purple and red curves in Fig. 4A). With increasing doping, the Dirac point of the TSSs is tuned from above to below $E_F$ and the Dirac-like dispersion is clearly observed [Figs. 4B (ii), 4C (ii and iii), and see fig. S9 for the momentum-dependent dispersions of the doped TDSSs]. These observations can be well reproduced by our *ab initio* calculations (Fig. 4D) by introducing a chemical potential shift on the surface. Furthermore, the direct observations of the whole TDSSs unambiguously demonstrate the nontrivial bulk topology. In addition to the significant energy shift of the Dirac-like TSSs (Fig. 4G), a less-dramatic downward shift of bulk bands is revealed (as highlighted by the red dashed curves in Figs. 4E and 4F). In particular, the electron-like VHS1 band around the $\bar{M}$ point increase in size and its bottom drops by about *30 meV*, as evidenced by the doping dependent energy distribution curve (EDC) taken at the $\bar{M}$ point (Fig. 4H, and see fig. S10 for the manipulation of the VHS1 band on the kagome termination). It is obvious that the distinct doping evolution of the Dirac bands (Fig. 4, C and G) and VHS1 band (Fig. 4H) cannot be explained by a simple rigid band shift of the $E_F$, reflecting the different chemical potential changes of the surface and bulk states with doping, as confirmed by the calculations (Fig. 4D).

**DISCUSSION**

Our ARPES measurements, combined with DFT calculations, reveal the desired $\mathbb{Z}_2$ TDSSs in a kagome metal, namely in the magnetic kagome metal GdV$_6$Sn$_6$. In addition, we identify multiple VHSs at the M point and demonstrate the successful manipulations of the TDSSs and VHSs by surface doping. With increasing doping, the Dirac point of the TDSSs is tuned from above to below $E_F$. Notably, upon sufficient doping, the lower branch of the TDSSs can merge into the bulk bands and only the upper branch crosses $E_F$, resulting in spin chirality reversal on the Fermi surfaces in GdV$_6$Sn$_6$ [as illustrated in the inset of Fig. 4 D (iii), for details see figs. S8 and S9]. Such highly tunable TDSSs show great promise for controlling spin current via local electrostatic gates [tuning the Dirac point below the $E_F$, see the inset of Fig. 4D (iii)], which deserves further experimental investigation. Moreover, the revealed tunability of the VHSs holds the potential for realizing exotic correlated states on the surface of GdV$_6$Sn$_6$ through Fermi surface nesting and sublattice interference by nonlocal interactions (27,30-33). On the other hand, as GdV$_6$Sn$_6$ tends to form magnetic ordering below the $T_m$, our results would stimulate future studies on the nontrivial states in magnetic phases. For instance, at low temperature, the magnetic Gd layers can become ferromagnetic by applying a weak magnetic field, which can gap out the surface Dirac cone and generate a nontrivial Chern number (see fig. S11 for the electronic structure of GdV$_6$Sn$_6$ in ferromagnetism). This realizes a quantum anomalous Hall state on the surface of GdV$_6$Sn$_6$ and the corresponding edge states and edge currents may be detected in a step edge in scanning tunneling microscopy (STM) measurements, or in thin-film transport measurements.

In summary, GdV$_6$Sn$_6$ is a novel magnetic kagome metal whose magnetic layer and kagome layer are separated. The crucial insights into the electronic structure, revealed by our work, provide compelling evidence that GdV$_6$Sn$_6$ is a long-sought kagome material that hosts TDSSs originating from a $\mathbb{Z}_2$ bulk topology. The presence of topologically nontrivial surface states combined with the ability to tune magnetic interactions in the magnetic layer makes GdV$_6$Sn$_6$ a promising candidate for the construction of topological devices. For example, a quantum anomalous Hall state can be induced on the surface in the ferromagnetic phase. Moreover, the multiple VHSs in the vicinity of the $E_F$ with their large density of states and tunability upon doping provide an ideal platform for correlated quantum states native to kagome lattices. Our results not only establish the kagome metal GdV$_6$Sn$_6$ as a fascinating playground for fundamental research connecting topological physics, electronic correlation and magnetism, but also open up a new avenue for the potential application of topological devices in spintronics, for which one could exploit the reversibility of the spin texture chirality of the surface states.

**MATERIALS AND METHODS**

**Sample growth** Single crystals of GdV$_6$Sn$_6$ were grown from the Sn-flux with the loading composition of Gd:V:Sn= 1:6:20. The sample was heated up to 1050 °C and stay at 1050 °C for 10 hours, then slowly cooled down to 650 °C at the speed of 3 °C/hour. The extra Sn flux was centrifuged at 650 °C. The grown single crystals were carefully examined by single crystal X-ray diffraction to obtain the accurate lattice parameters and atomic coordinates. Accordingly, GdV$_6$Sn$_6$ crystallizes in a layered structure with the space group P6/mmm.

**ARPES measurements** The GdV$_6$Sn$_6$ samples were cleaved *in-situ* with a base pressure of better than $5 \times 10^{-11}$ torr. Regular angle-resolved photoemission (ARPES) measurements were performed at the ULTRA endstation of the Surface/Interface Spectroscopy (SIS) beamline of the Swiss Light Source using a Scienta-Omicron DA30L analyzer. The temperature was 20 *K*, and total energy resolution was better than 15 *meV*. The Fermi level was determined by measuring a polycrystalline Au in electrical contact with the samples. The spin-resolved ARPES experiments were conducted at the low-energy branch of the APE beamline (APE-LE) of the Elettra synchrotron (Trieste, Italy), equipped with a DA30 analyzer combined with VLEED spin detectors.

**DFT calculations** Band structure calculations were performed by using the method of first-principles density functional theory (DFT) as implemented in the QUANTUM ESPRESSO (QE) code (34). The cutoff energy for expanding the wave functions into a planewave basis was set to 60 Ry and the adopted K-point grid is 9 x 9 x 5. The exchange correlation energy was described by the generalized gradient approximation (GGA) using the PBE functional (35). The calculations were done for nonmagnetic GdV$_6$Sn$_6$ with spin-orbit coupling. We used the maximally localized Wannier functions (MLWFs) to construct a tight-binding model by fitting the DFT band structure, where 118 MLWFs were included (Gd *d*; V *d*; Sn *s*, *p*;) (36) and then we used the surface state Green's function method to calculate topological surface states (37). To simulate the doping evolution of the band structure, we model the potassium deposition by introducing a chemical potential shift only in the top surface layer. For all calculations, we used the experimentally determined crystal structure and lattice constants (a = 5.5348 Å and c = 9.1797 Å).

**Acknowledgements:** The authors wish to thank I. Vobornik and J. Fuji for their technical assistance in spin-ARPES measurements. The work at PSI was supported by the Swiss National Science Foundation under Grant. No. 200021-188413, the Sino-Swiss Science and Technology Cooperation (Grant No. IZLCZ2-170075), and the NCCR MARVEL, a National Centre of Competence in Research, funded by the Swiss National Science Foundation (grant number 182892). Y.H. was supported by the National Natural Science Foundation of China (12004363). J.Z.M. was funded by the National Natural Science Foundation of China (12104379), Guangdong Basic and Applied Basic Research Foundation (2021B1515130007), and City University of Hong Kong (project number 7020067, 9610489, 9680339). The work at Rutgers was supported by U.S. DOE-BES under Contract DE-SC0022156. **Author Contributions:** Y.H. and J.Z.M. designed the research. Y.Y. grew and characterized the crystals with guidance from W.X.. X.W. performed the theoretical calculations with the support from A.P.S.. Y.H. and J.Z.M. performed the regular ARPES experiments with help from N.C.P. and M.S.. Y.H. and S.G. performed the Spin-resolved ARPES experiments. Y.H. analyzed the data and discussed with J.Z.M. and M.S.. Y.H. draw the figures with help from X.W. and J.Z.M.. Y.H. wrote the paper with inputs from J.Z.M. and X.W.. All authors participated in discussions and comments on the paper. M.S., J.Z.M. and Y.H. supervised the project.

**Competing interests:** The authors declare that they have no competing interests. **Data and materials availability:** All data needed to evaluate the conclusions in the paper are present in the paper and/or the Supplementary Materials. Additional data related to this paper are available at a public repository (MARVEL Materials Cloud Archive) with the same title of this paper (https://archive.materialscloud.org).

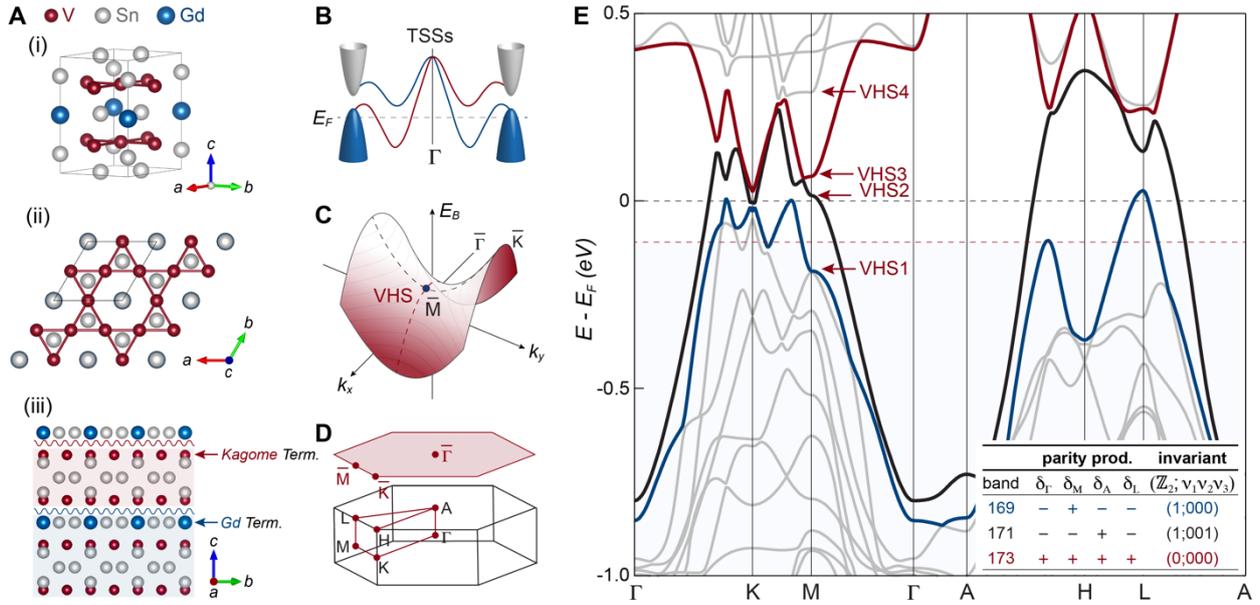

**Fig. 1. Crystal structure, topological classification, and van Hove singularities in kagome metals GdV$_6$Sn$_6$.** (**A**) Crystal structure of GdV$_6$Sn$_6$ showing the unit cell (i), top view looking along the *c* axis and showing the V kagome plane (ii), side view showing two possible surface terminations as indicated by wavy line (iii). (**B**) Sketch of topologically nontrivial states (TSSs) in GdV$_6$Sn$_6$. Cones and curves represent bulk states and TSSs, respectively. (**C**) Schematic of a van Hove singularity (VHS) in a two-dimensional electron system. (**D**) Bulk Brillouin zone (BZ) of GdV$_6$Sn$_6$ and the projection of the (001) surface BZ, with high symmetry points marked. (**E**) Density functional theory calculated electronic structure of GdV$_6$Sn$_6$. The red dashed line indicates the Fermi level suggested by the ARPES measurement. The inset shows the parity products classifying the $\mathbb{Z}_2$ invariant for each band. Bands 169 (blue) and 171 (black) are characterized by a strong topological invariant, $\mathbb{Z}_2$ =1, while band 173 (red) is trivial with no topological invariants.

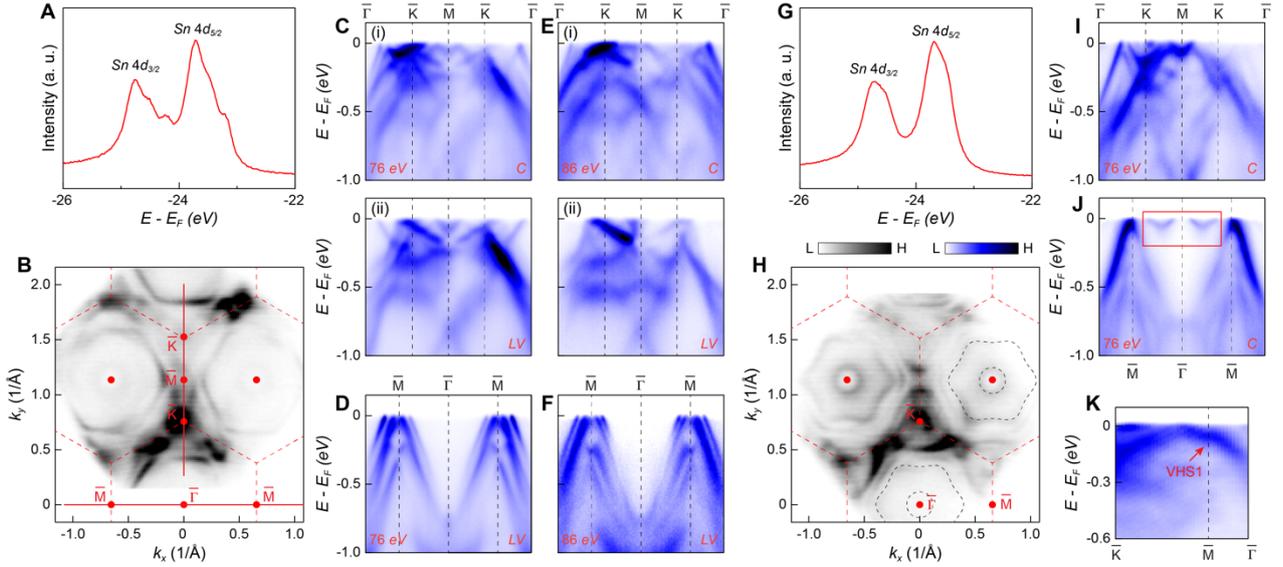

**Fig. 2. Termination dependence of the electronic structure in GdV$_6$Sn$_6$.** (**A**) XPS spectrum of *in-situ* freshly cleaved GdV$_6$Sn$_6$, from which we suggest that the surface termination is the kagome layer (fig. S3). (**B**) Fermi surface mapping measured on the kagome termination. The BZ is marked with the red dashed hexagon. (**C**) ARPES spectrum taken along the $\bar{\Gamma}$ - $\bar{K}$ direction on the kagome termination, measured with 76 *eV* circular (*C*) (i) and linear vertical (*LV*) (ii) polarized light. The momentum path is indicated by the red solid line in (B). **D** Same as (C), but taken along the $\bar{\Gamma}$ - $\bar{M}$ direction, and measured with *LV* polarization. (**E** and **F**) Same as (C and D), but probed with 86 *eV* photons. (**G** and **H**) Same as (A, B), but measured on the Gd termination. The black dashed curve marks the electronic pocket. (**I** and **J**) Same as (C, D), but measured with *C* polarization on the Gd termination. (**K**) Band dispersion along the $\bar{K}$ - $\bar{M}$ - $\bar{\Gamma}$ path showing the VHS1 at the $\bar{M}$ point.

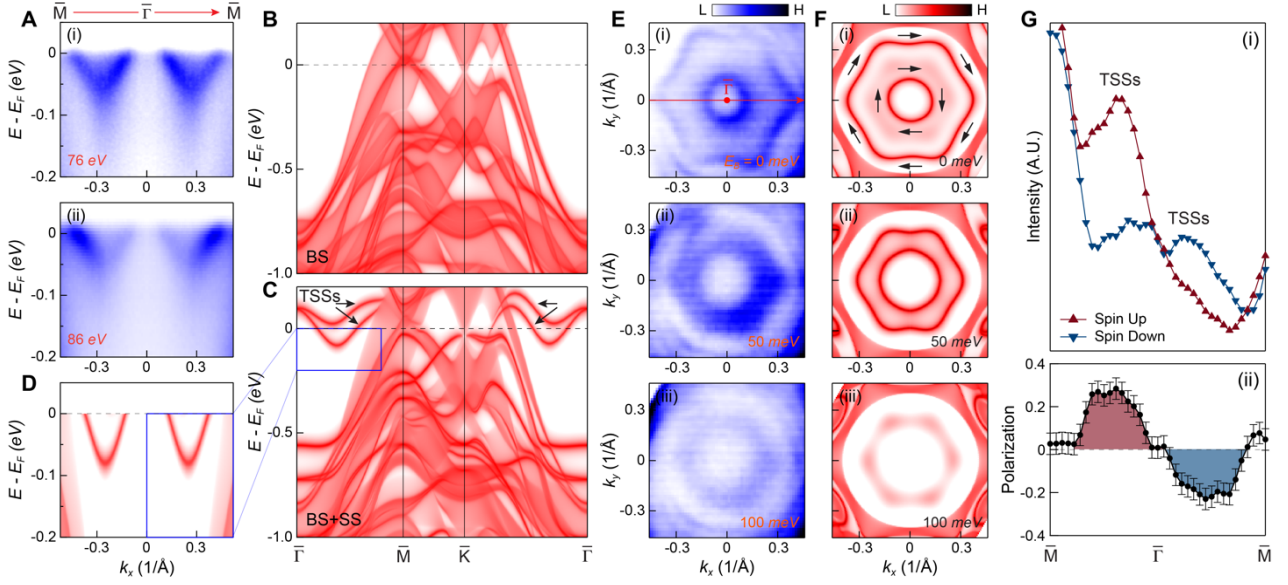

**Fig. 3. $\mathbb{Z}_2$ Topological surface states in GdV$_6$Sn$_6$.** (**A**) Photon energy-dependent ARPES spectra taken along the $\bar{\Gamma}$ - $\bar{M}$ direction, measured with 76 *eV* (i) and 86 *eV* (ii). (**B** and **C**) The (001) surface Green's function projection of pure bulk states (BS)(B) and the states [BS and surface states (SS)] on Gd termination (C). (**D**) Zoom-in plot of the calculated TSSs with the same energy-momentum range as the experimental dispersions in (A). (**E** and **F**) Side-by-side comparison between experiments (E) and calculations (F), which exhibits excellent agreement, of three representative constant-energy contours ($E_B$ = 0, 50, 100 *meV*). (**G**) Spin-resolved momentum distribution curve (MDC), collected along the orange line in fig. S8(a). The red and blue symbols in (i) are the intensity of the spin-up and spin-down states, respectively. The black curve indicates the spin polarization (ii). The spin texture of TSSs on Fermi surface are indicated by the black arrows in [F(i)] (for details see fig. S8).

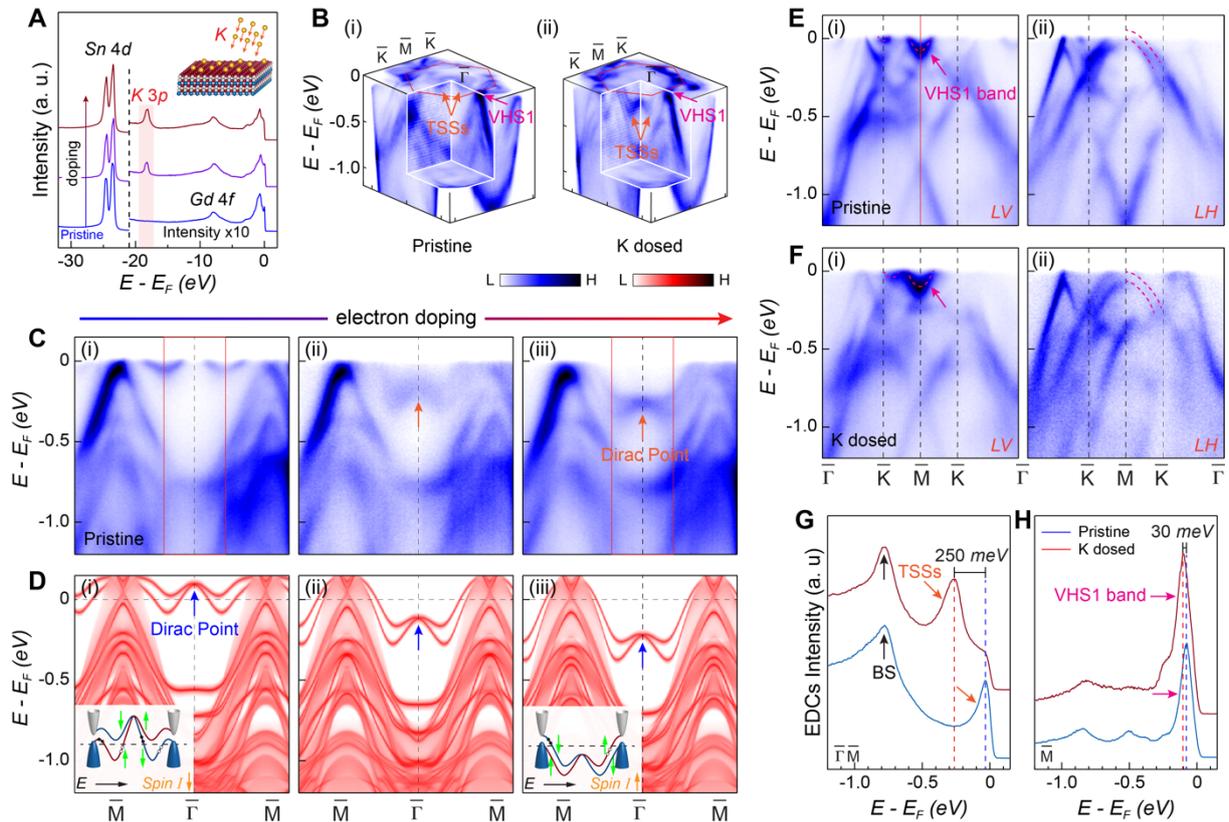

**Fig. 4. Manipulation of the TSSs and VHS via *in-situ* potassium deposition.** (**A**) Doping dependence of the XPS spectrum recorded on the Gd termination, showing the characteristic Sn 4*d* and Gd 4*f* peaks. Upon potassium deposition, the K 3p peak emerges (purple and red curves), which is absent on the pristine surface (blue curve). (**B**) 3D intensity plot of the electronic structure measured on the pristine (i) and K dosed surfaces (ii). The orange arrow highlights the TSSs. (**C** and **D**) Doping evolution of the band structure along the $\overline{\Gamma}$ - $\overline{M}$ direction from experiments (C) and calculations (D). The Dirac cone of the TSSs is above the Fermi level ($E_F$) before doping (pristine, i) and is tuned below $E_F$ with electron doping (ii, iii). The arrows indicate the Dirac cone of the TSSs. The insets in [D(i)] and [D(iii)] show the schematic of distinct spin current of TDSSs without and with doping. (**E**) ARPES spectra taken along the $\overline{\Gamma}$ - $\overline{K}$ direction, on the pristine surface (Gd termination), measured with 76 *eV LV* (i) and linear horizontal (*LH*) (ii) polarizations. Red dashed curve highlights bulk bands. The arrow indicates the VHS1 band. (**F**) Same as (E), but measured on the K dosed surface. (**G**) Doping evolution of the integrated energy distribution curve (EDC) taken around $\overline{\Gamma}$. The integration window of the EDC is represented by the red box in [C(i)] and [C(iii)]. Red and black arrows mark the TSSs and BS, respectively. (**H**) Doping evolution of the EDC extracted at the $\overline{M}$ point, as indicated by the red line in [E(i)], measured with *LV* polarization.